# Four-Dimensional-Spacetime Atomistic Artificial Intelligence Models

Fuchun Ge[1], Lina Zhang[1], Yi-Fan Hou[1], Yuxinxin Chen[1], Arif Ullah[2], Pavlo O. Dral[1*]

[1]*State Key Laboratory of Physical Chemistry of Solid Surfaces, College of Chemistry and Chemical Engineering, Fujian Provincial Key Laboratory of Theoretical and Computational Chemistry, and Innovation Laboratory for Sciences and Technologies of Energy Materials of Fujian Province (IKKEM), Xiamen University, Xiamen, Fujian 361005, China*

[2]*School of Physics and Optoelectronic Engineering, Anhui University, Hefei 230601, China*

E-Mail: dral@xmu.edu.cn

**Abstract**

We demonstrate that AI can learn atomistic systems in the four-dimensional (4D) spacetime. For this, we introduce the 4D-spacetime GICnet model which for the given initial conditions – nuclear positions and velocities at time zero – can predict nuclear positions and velocities as a continuous function of time up to the distant future. Such models of molecules can be unrolled in the time dimension to yield long-time high-resolution molecular dynamics trajectories with high efficiency and accuracy. 4D-spacetime models can make predictions for different times in any order and do not need a stepwise evaluation of forces and integration of the equations of motions at discretized time steps, which is a major advance over the traditional, cost-inefficient molecular dynamics. These models can be used to speed up dynamics, simulate vibrational spectra, and obtain deeper insight into nuclear motions as we demonstrate for a series of organic molecules.

**TOC Graphic**

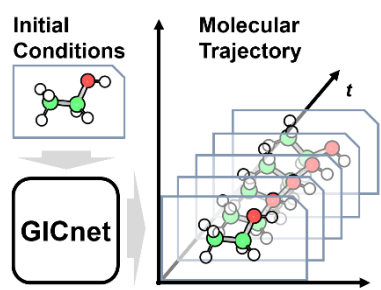

The matter around us consists of interacting atoms, which are not static but always in motion subject to the laws of quantum mechanics (QM).[1, 2] Unfortunately, atomistic modeling beyond static 3D structures with the first-principles QM approaches is computationally extremely costly even for small systems.[3-5] An approximated approach for modeling temporal changes of nuclear positions is molecular dynamics (MD) which propagates nuclei classically and sequentially, one step at a time, and requires the information about forces at each time step for integration of Newton's equations of motion.[6-9] Despite approximations made in MD, it is an extremely important tool in research and development with almost a million publications in Web of Science, with the annual number of publications steadily growing and amounting to over 50 thousand in recent years. Because of MD's sequential nature and the need for evaluating forces and integration at each time step, MD is however inherently a slow approach, whose accuracy and speed strongly depend on many factors such as a method for calculating forces and the chosen length of a time step $\Delta t$. This length generally should not be too short (because it would require too many steps to reach the final simulation time) or too long (because the quality of the trajectory would deteriorate), and its choice influences the outcome of trajectory propagation (Figure 1a).[10-12] Another point is that in practice MD is discretized rather than continuous.

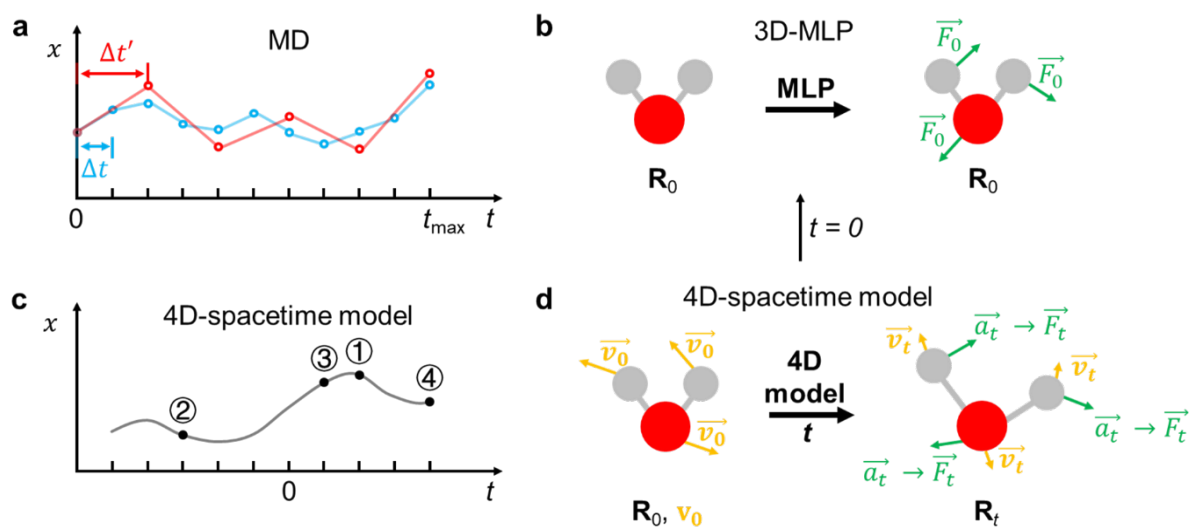

**Figure 1. Comparison of 4D-spacetime atomistic artificial intelligence models with traditional molecular dynamics (MD) approach. a**. MD can be propagated up to a limited time and with finite time steps influencing the trajectory (i.e., outcome is different for two different time steps $\Delta t$, blue, and $\Delta t'$, red). **b**. 3D-machine learning potentials (3D-MLP) only predict forces for a current geometry to propagate dynamics. **c**. 4D-spacetime models predict atomic positions at any time in any order, e.g., in the order 1 → 2 → 3 → 4 for points shown in the figure. **d**. 4D-spacetime atomistic artificial intelligence

models can predict new geometry, with velocities and forces for each atom at different times, making 3D-MLP a special case of 4D-spacetime models. $x$ is any property propagated by trajectory, **R** – molecular geometry, $t$ – time, $t_{\max}$ – trajectory propagation time, $v$ – velocity (orange), $a$ – acceleration and $F$ – force (green), subscript 0 denotes initial conditions and arrows above letters denote vectors.

The forces can be calculated with QM electronic structure methods within Born–Oppenheimer (BO) approximation as in BOMD variant of MD[13-16], but BOMD is very slow. A fast alternative is molecular mechanics (MM) MD, but MM methods are generally less transferable.[17] Thus, many artificial intelligence (AI)-based approaches were suggested to reduce the cost of Born–Oppenheimer dynamics while retaining their accuracy via training machine learning (ML) potentials (MLPs) to serve as the surrogate models of QM approaches.[18-20] These MLPs are still 3D models, which are functions $E^{\text{pot}} = f(\mathbf{R})$ predicting potential energies $E^{\text{pot}}$ and forces ($F_{br} = -\partial E^{\text{pot}}/\partial R_{br}$ for atom $b$ and coordinate $r$) for the 3D static structure **R** at each MD time step (Figure 1b) propagated in exactly the same way as BOMD or MM MD. However, 3D-MLP MD inherits the same fundamental problem of MD: it is sequential, slow, needs integration, and depends on the length of the time step. Accelerating simulations can be only done via optimizing algorithms and using expensive, vast computational resources.[5]

Here we propose a radically different concept to model matter in 4D spacetime at an atomistic level using AI. Our ultimate goal is to create such a 4D-spacetime atomistic model $f$ which can predict atomistic structure in 3D space defined by nuclear coordinates $\mathbf{R}_t$ at a given time $t$ in the near and distant future. As input, this model only needs an initial structure $\mathbf{R}_0$ and velocity $\mathbf{v}_0$ (initial conditions) at time zero:

$$\mathbf{R}_t = f(\mathbf{R}_0, \mathbf{v}_0, t). \qquad (1)$$

4D-spacetime models are by definition functions of continuous time, do not require arbitrary time steps, do not need to be evaluated sequentially one-step-at-a-time, and completely eliminate the integration of Newton's equations of motion at each time step. 4D-spacetime model predictions do not depend recursively on any previous time step, i.e., predictions can be made independently for different times (Figure 1c). This all makes such models substantially more computationally efficient than any traditional 3D MD approach while they can be used, wherever MD is used,[21-27] e.g., for simulating physicochemical transformation rates and spectra.

Once a 4D-spacetime model is constructed, one can obtain from it many other properties in a straightforward manner. For example, at any time $t$, the first- and second-order time derivatives of positions would give us the velocity $v_{br,t}$ and acceleration $a_{br,t}$ of an atom $b$ and coordinate $r$:

$$v_{br,t} = \frac{dR_{br,t}}{dt} = \frac{d\left(f(\mathbf{R}_0, \mathbf{v}_0, t)\right)_{br}}{dt}, \qquad (2)$$

$$a_{br,t} = \frac{dv_{br,t}}{dt} = \frac{d^2\left(f(\mathbf{R}_0, \mathbf{v}_0, t)\right)_{br}}{dt^2}. \qquad (3)$$

In turn, from acceleration, we can calculate forces for atoms with masses $m$:

$$F_{br,t} = m_b a_{br,t}. \qquad (4)$$

Thus, 3D-MLP approaches may be considered as a special case of 4D-spacetime models, applied only for static structures (Figure 1d).

In this work, we introduce both the general concept of 4D-spacetime atomistic artificial intelligence models as well as our specific realization of this concept – the GICnet model which is shown to be able to efficiently and accurately represent polyatomic molecules. In the following, we will demonstrate how to construct the GICnet models and apply them instead of traditional, stepwise MD for propagating long-time trajectories to simulate conformational distributions and rovibrational spectra. We will also show how the 4D-spacetime models can be used to obtain unique insight into nuclear motions in molecules.

**Model design.** The development of machine learning potentials is an active research field and numerous studies indicate that even accurately learning a single property such as potential energy as a function of nuclear coordinates is nontrivial. Designing a 4D-spacetime model which is able to accurately learn the time-evolution of many properties, i.e., positions of all nuclei, is an expectedly even more challenging task. We can create 4D-spacetime models which essentially ideally predict their geometries at any time in the past and future only for the simplest systems such as diatomic molecules with one vibrational degree of freedom leading to periodic dynamical behavior (see the Supporting Information (SI) for a hydrogen molecule example).

The realistic polyatomic molecules of chemical interest are, however, very complex objects in 4D spacetime, for which we have developed the GICnet model able to predict the time evolution of nuclear positions up to a distant but finite future. GICnet is based on deep neural networks (NNs, see the SI) which are universal approximators[28] and were proven to be successful in related applications ranging from MLPs[19, 20] to learning trajectories of non-atomistic objects (trajectories from model Hamiltonians and potentials),[29-33] increasing integration time steps,[34] improving time resolution in dynamics,[35] learning partial differential equations,[36-38] analyzing dynamics,[39] enhancing sampling,[40] and predicting dynamics outcomes[41].

We choose to learn nuclear positions **R** in redundant internal coordinates (IC, see the SI) which provide an intuitive representation of chemical structures. ICs are widely used in QM simulations and known[42] to be reasonably good for 3D-MLP models. This also allows us to enforce the conservation of momentum when we use the GICnet model for dynamics propagation by reconstructing the Cartesian coordinates of nuclear positions from their internal coordinates. Investigation into the design of better representations is expected to be a challenging but potentially very rewarding if successful.

Another requirement of the model is to make sure that the GICnet at time zero exactly recovers initial conditions. Thus, we design such a function $f(\mathbf{R}_0, \mathbf{v}_0, t)$ in Eq. 1 to obtain $\mathbf{R}_0$ and $\mathbf{v}_0$ at $t = 0$:

$$\mathbf{R}_t = f(\mathbf{R}_0, \mathbf{v}_0, t) = \mathbf{R}_0 + \mathbf{v}_0 t + f_{\text{damp}}(t) f_{\text{NN}}(\mathbf{R}_0, \mathbf{v}_0, t), \qquad (5)$$

$$\mathbf{v}_t = \frac{\partial \mathbf{R}_t}{\partial t} = \mathbf{v}_0 + f_{\text{damp}}(t) \frac{\partial f_{\text{NN}}(\mathbf{R}_0, \mathbf{v}_0, t)}{\partial t} + \frac{df_{\text{damp}}(t)}{dt} f_{\text{NN}}(\mathbf{R}_0, \mathbf{v}_0, t), \qquad (6)$$

where $f_{\text{NN}}(\mathbf{R}_0, \mathbf{v}_0, t)$ is an NN function taking as an input time $t$, IC $\mathbf{R}_0$ and their time derivatives $\mathbf{v}_0$ at time zero, and $f_{\text{damp}}(t)$ is a damping function that should satisfy conditions that it and its time derivatives at $t = 0$ are zero, i.e., $f_{\text{damp}}(0) = 0$ and $df_{\text{damp}}(t)/dt = 0$ (see the SI for the choice of both $f_{\text{NN}}$ and $f_{\text{damp}}$).

As with any AI model, GICnet requires training data, which should contain the set of nuclear position $\mathbf{R}_t$ at different times $t$. This data can be generated by running traditional MD trajectories from different initial conditions and the challenge is that the training data should capture the required phase space. We cannot learn infinitely-long MD, hence, we choose the maximum time $t_c$ (cutoff time) which GICnet attempts to learn. In order to generate reasonable set of labeled points to train GICnet, we propagate dozens of medium-length traditional MD trajectories which we split into the time segments $[0, t_c]$ to generate millions of short trajectories (Figure 2a). We discuss the influence of the segment length later.

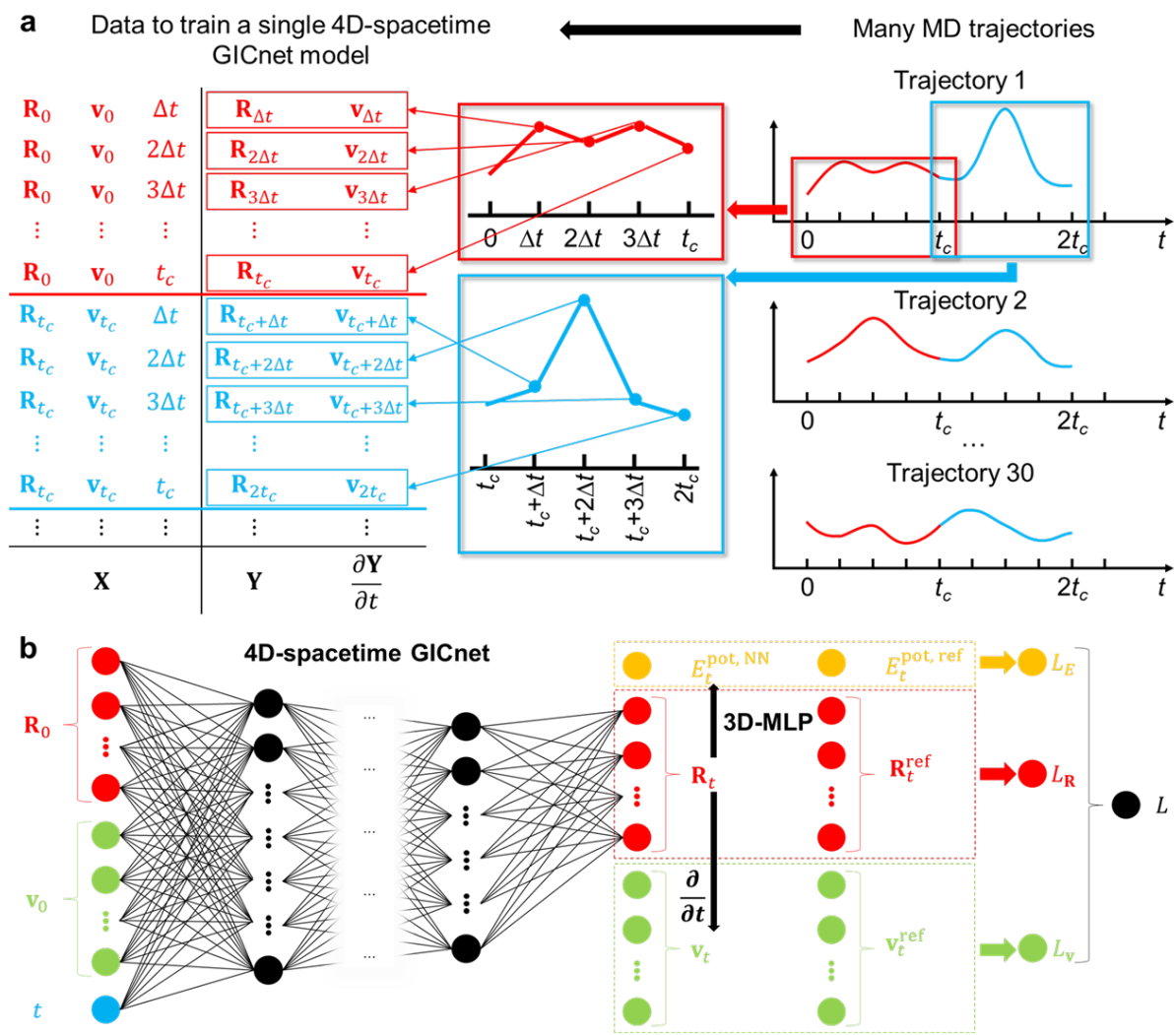

**Figure 2. Transforming traditional molecular dynamics (MD) trajectories into labeled data for learning with a neural network (NN). a**. Transforming many medium-length reference MD trajectories into short segments $[0, t_c]$ and eventually to the labeled data for training NNs. Different colors correspond to trajectory segments of the same length. **b**. Training GICnet model. **X** – input for NN, **Y** – labels to learn, $t_c$ – time cutoff, $\Delta t$ – MD time step, **R** – nuclear positions, **v** – nuclear velocities, subscripts of **R** and **v** indicate the time step they are taken from, superscript – either NN predictions or reference values, $L$, $L_\mathbf{R}$, $L_\mathbf{v}$, $L_E$ denote total loss, loss for geometries, velocities, and energies, respectively.

In order to improve the accuracy of the GICnet models we train them on both nuclear positions and their time derivatives. However, we discovered that even though the training and validation errors are small with such models, they could not produce stable long trajectories as the molecules very quickly either implode or explode. Thus, we found that it is crucial to regularize NNs to make sure that the geometries it produces have reasonable potential energies. For this, during training of GICnet we evaluate potential energies at the geometries it produces with the auxiliary 3D-MLP trained on one million of the potential energies from the training data (feed-forward fully-connected NN taking IC as input, see the SI). Alternatively, the GICnet architecture could be modified to also predict potential energies but currently it is unclear whether such modifications would be better than our approach of training the separate auxiliary 3D-MLP model.

Our final loss function $L$ used for training the GICnet model includes weighted terms for nuclear coordinates, their time derivatives, and potential energies (Figure 2b):

$$L = L_\mathbf{R} + L_\mathbf{v} + L_E = w_B L_B + w_A L_A + w_D L_D + w_{vB} L_{vB} + w_{vA} L_{vA} + w_{vD} L_{vD} + w_{E_\mathrm{pot}} L_{E_\mathrm{pot}}, \quad (7)$$

where $w$ and $L$ are weights and losses. $L_\mathbf{R}$, $L_\mathbf{v}$, $L_E$ denote loss for geometries, velocities, and energies, respectively. Subscripts B, A, D, and $E_\mathrm{pot}$ denote bond lengths, bond angles, dihedral angles, and potential energies, respectively, and the leading $v$ in subscriptions indicates the loss corresponding to the respective time derivatives (SI). Loss in energies is evaluated with the help of the auxiliary 3D-MLP model.

**Trajectory propagation.** After training, the model can be used for very fast predictions of MD trajectories for new unseen initial conditions and they can be only done for a time segment $[0, t_\mathrm{segm}]$ with $t_\mathrm{segm} \leq t_c$, i.e., only up to a distant, but finite time $t_c$ used for training the GICnet model. For the fastest simulations, we set $t_\mathrm{segm} = t_c$ in all trajectories and make predictions for equally spaced times with distance 0.05 fs within the segment to ensure high resolution. To enable long MD, the GICnet takes the last nuclear positions and velocities at the end of the previous time segment $[0, t_\mathrm{segm}]$ as new initial conditions and predicts the next time segment to obtain a long-time MD trajectory (see Figure 3a and the SI for details). While the propagation with segments might seem analogous to the traditional MD where propagation is also done with time steps, there are crucial differences: 1) the concept of 4D-spacetime models is not limited to use time segments and for simplest molecules one could predict the nuclear configuration up to arbitrary time (see SI for an example of hydrogen), 2) for larger molecules, the time segments are still much longer than time steps (at least an order of magnitude) and the accuracy does not drop as dramatically with longer time segment as the quality of MD when time step is increased (see results and discussion below), 3) the predictions within time segment can be done for any arbitrary times, while time steps are discrete, 4) one can extract chemical insights directly from the 4D-spacetime model while in case of MD one would need to propagate and *a posteriori* analyze trajectories (see also below). Another seemingly analogous approaches are NN models specifically designed for learning sequential information: the 4D-spacetime models are different in that they learn sequential information not as a sequence but as a function of time, thus, even time segments can be of arbitrary, adjustable length (we use the same length for simplicity). In addition, our related study on learning the sequential information in the dissipative quantum dynamics revealed that the specialized NN models are not necessarily better choices than simpler models such as feed-forward NN used in the GICnet.[43]

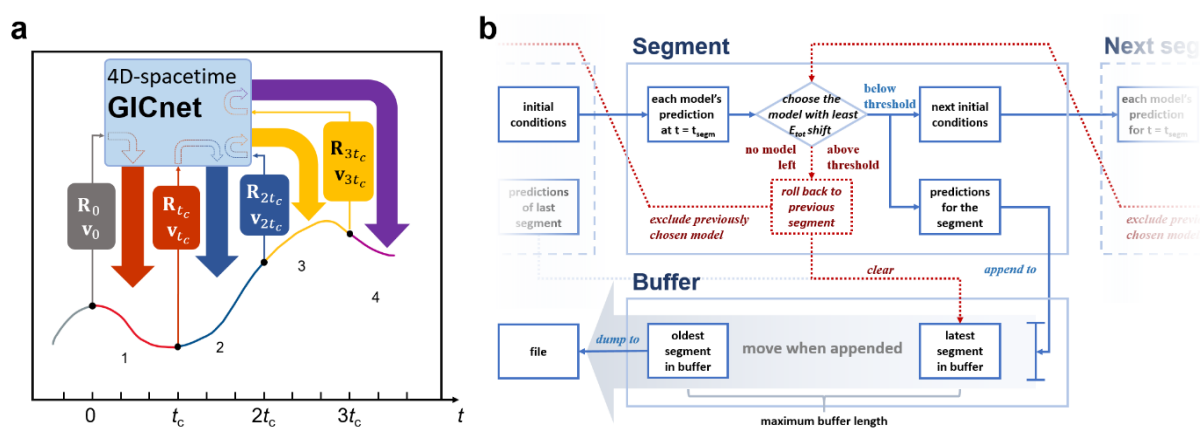

**Figure 3. Propagating long molecular dynamics (MD) trajectories with the 4D-spacetime GICnet model.** The GICnet model is used to make predictions only within segments of lengths $t_c$ and the end-points of the previous segment are used as initial conditions for the next segment, highlighted by different colors. **R** – nuclear positions, **v** – nuclear velocities. Potential energies required for calculating the total energies are evaluated with the help of the auxiliary 3D-MLP model.

Long-time trajectory propagation is a challenge even for traditional MD, especially propagated with 3D-MLPs[44, 45], because the molecules sometimes unphysically explode as we also observed in many reference trajectories during the preparation of the training set (unphysical explosions happen because the 3D-MLP does not describe a dissociation part of PES correctly by severely underestimating the potential energies;[46] we removed such exploding reference trajectories and did not use them for training, see the SI). Thus, when we trained the preliminary versions of the GICnet models without physical constraints introduced below, we observed that the molecules may explode during long-time MD. We attribute such instabilities to substantial violation of the conservation of energy, even though the total energy fluctuations in the 4D-spacetime and traditional MD have very different effects on trajectories: the errors in total energy in traditional MD quickly accumulate in each time step while in 4D-spacetime MD they are quite independent for different times as the nuclear positions are predicted as a function of time, and the wrong total energy in one time does not necessarily lead to the worse prediction of nuclear positions in the later time.

To increase the stability of the GICnet long MD trajectories, we introduce a strategy to decrease the drifts in the total energy during MD (Figure 3b). Since the 4D-spacetime GICnet models are not potential-energy functions, but predict nuclear configurations and velocities, it does not directly yield total energies and can only be used to calculate kinetic energies. In order to approximate total energies, we need to estimate the potential energies with the auxiliary 3D-MLP model used to regularize the GICnet model during its training (see above). We train an ensemble of eight GICnet models for a molecule and choose the model with the smallest drift in total energy at the end of each time segment; if the drift is still too large, we roll back to previous segments and switch to other models to decrease the drift at the end of several consecutive time segments (see the SI). In our test cases for different molecules (see below), the estimated total energy during long-time MD has little systematic drift and its fluctuations are tolerable (see the SI). Note that reported[47, 48] drifts in total energy during MD simulations even with the pure QM methods can be rather large (up to ca. 100 kcal/mol) and, although they are highly undesirable and should be, in principle, avoided, they still could be tolerated in the published literature and were shown in some cases not to affect the MD outcome much.[47]

Our mitigation strategy using the above physical constraints on energy conservation ensures that the GICnet trajectories stay stable for long-time propagation, e.g., for one of our test molecules (ethanol, see below), propagation is stable at least up to 10 ns – the longest time we tested. Such stability is quite an achievement for a pure AI model.[44, 45] In the future, the stability of the models may be improved via automatic construction of the data sets with active learning (similar to common approaches[49, 50] developed for 3D-MLP).

**Dynamics of polyatomic molecules.** Here we analyze the performance of 4D-spacetime GICnet models for ten test molecules (ethanol, benzene, uracil, naphthalene, aspirin, salicylic acid, malonaldehyde, toluene, paracetamol, and azobenzene) from the MD17 data set. The training trajectories were obtained with the universal ANI-1ccx[51] 3D-MLP potential approaching gold-standard coupled-cluster accuracy (see the SI). We first evaluate the GICnet models for predictions within the time segments of length $t_{\text{segm}} = t_c$ in the test trajectory also propagated with ANI-1ccx (labeled test data generated as shown in Figure 2a). The test mean root-mean-squared-errors (mean RMSEs) for bond lengths, bond angles, and dihedral angles are rather small (Figure 4), e.g., for bond lengths they range from 0.0025 Å for uracil ($t_c$ = 10 fs) and up to 0.035 Å in the worst case of paracetamol and $t_c$ = 20 fs. From the plots, it is obvious that attempting to learn longer time is more challenging as mean RMSEs are larger for $t_c$ = 20 fs compared to $t_c$ = 10 fs. In general, we found that for the test molecules, the sweet spot for the time segment length is 10–20 fs which strikes the best balance between model efficiency (the longer the time segment the faster it is) and accuracy (longer time segments are more difficult to learn while we did not attempt to optimize the models for shorter time segments). Additional analysis of the influence of time cutoff on an example of ethanol is given in the SI.

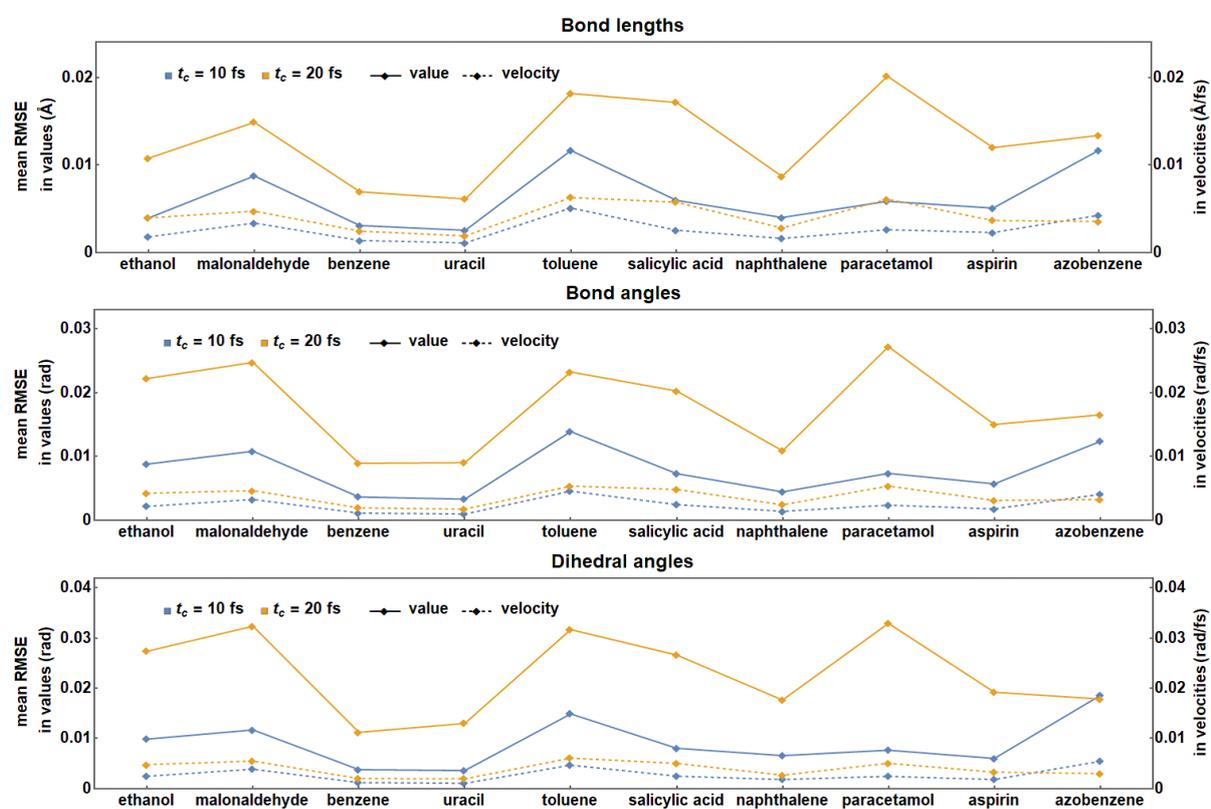

**Figure 4. Test errors in geometries and velocities.** The test root-mean-squared errors (RMSEs) were evaluated with the GICnet models for each bond length and angle and averaged to yield mean RMSEs. Results are shown for different test time segments $[0, t_c]$ where $t_c$ is time cutoff.

In order to evaluate the GICnet performance for long-time MD, we propagate test trajectories with GICnet and the reference 3D-MLP potential ANI-1ccx. For a brief overview of the trajectory qualities, we plot the RMSD (root-mean-squared deviation) between the nuclear coordinates in the reference and GICnet trajectories as a function of time up to 1 ps for each molecule (Figure 5). In general, the RMSD plots show reasonable agreement up to a rather long time. However, they also show that the behavior of the GICnet trajectories compared to the reference ones strongly depends on the molecule, e.g., if we look at the diagonals directly comparing the RMSD between GICnet and reference structures for the same time, deviations for some of the molecules like benzene stay very small for a much longer period than for others like ethanol. If one focuses on the x- and y-axis of the RMSD plots, one can observe how often similar structures repeat themselves and when the trajectory enters a new unexplored region. From this perspective, there are few molecules (e.g., azobenzene, paracetamol, and aspirin), when either GICnet or reference trajectories enter the new region sooner – it means that the shown 1 ps period is not long enough to explore all the regions. For a more complete comparison, MDs with both GICnet and ANI-1ccx are provided in Supporting Video S1.

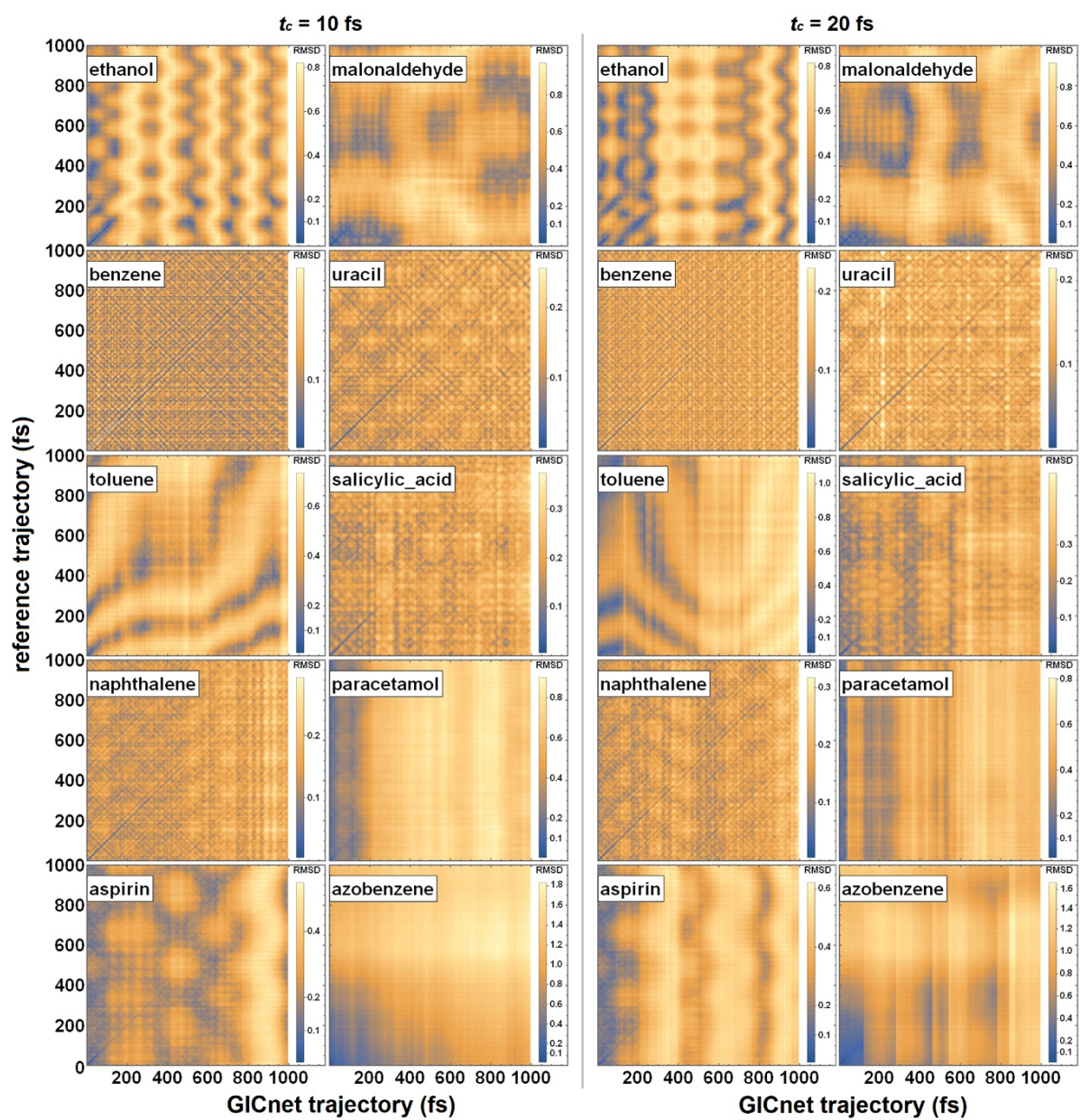

**Figure 5. Root-mean-squared deviation (RMSD) plots comparing the reference ANI-1ccx and 4D-spacetime GICnet trajectories.** RMSD plots for tested molecules are shown for different time cutoffs ($t_c$): 10 fs (left), and 20 fs (right).

To better judge the quality of the long MD trajectories, we also generate the rovibrational (power) spectra obtained through Fourier transform[52] of the test trajectories propagated longer than 1 ps (Figure 6). Similar spectra comparisons were used to gauge the quality of the MD propagated with the 3D-MLP.[50] Spectra were obtained from trajectories propagated for 50 ps in the cases of ethanol, malonaldehyde, benzene, and paracetamol, for 20 ps for uracil, and for 10 ps for other molecules (depending on how long it takes to converge the 4D-spacetime

GICnet power spectrum for each molecule, see the SI). We use the 20-fs GICnet models for benzene and uracil and 10-fs models for all other molecules because these two molecules have mean RMSEs for $t_c = 20$ fs similar to mean RMSEs of other molecules for $t_c = 10$ fs (Figure 4). In the SI, we report the power spectra for both time cutoffs of 10 and 20 fs for all molecules.

As with any AI model, we cannot expect it to have perfect accuracy for the tests, thus, it is important to put our analysis in the context of more established AI models. Thus, we additionally calculate power spectra with the traditional dynamics propagated with a popular 3D-MLP DeepPot-SE trained on data from the ANI-1ccx trajectories (SI). All power spectra are shown in Figure 6 for ten molecules and the similarities between GICnet and ANI-1ccx spectra are quite good and comparable to similarities between DeepPot-SE and ANI-1ccx spectra: for nine out of ten molecules, GICnet's similarity is close to or above 0.7, for four – even above 0.9 (spectra similarities calculated as described in literature[53]). It is noteworthy that for four out of ten molecules, the GICnet trajectories were run up to 50 ps while the reference trajectories used to generate the training data were run up to a much shorter 10 ps which shows that our approach is stable. Thus, the GICnet models can be useful to save computational time when it is necessary to run long trajectories while the training can be done on much less computationally expensive relatively short reference trajectories.

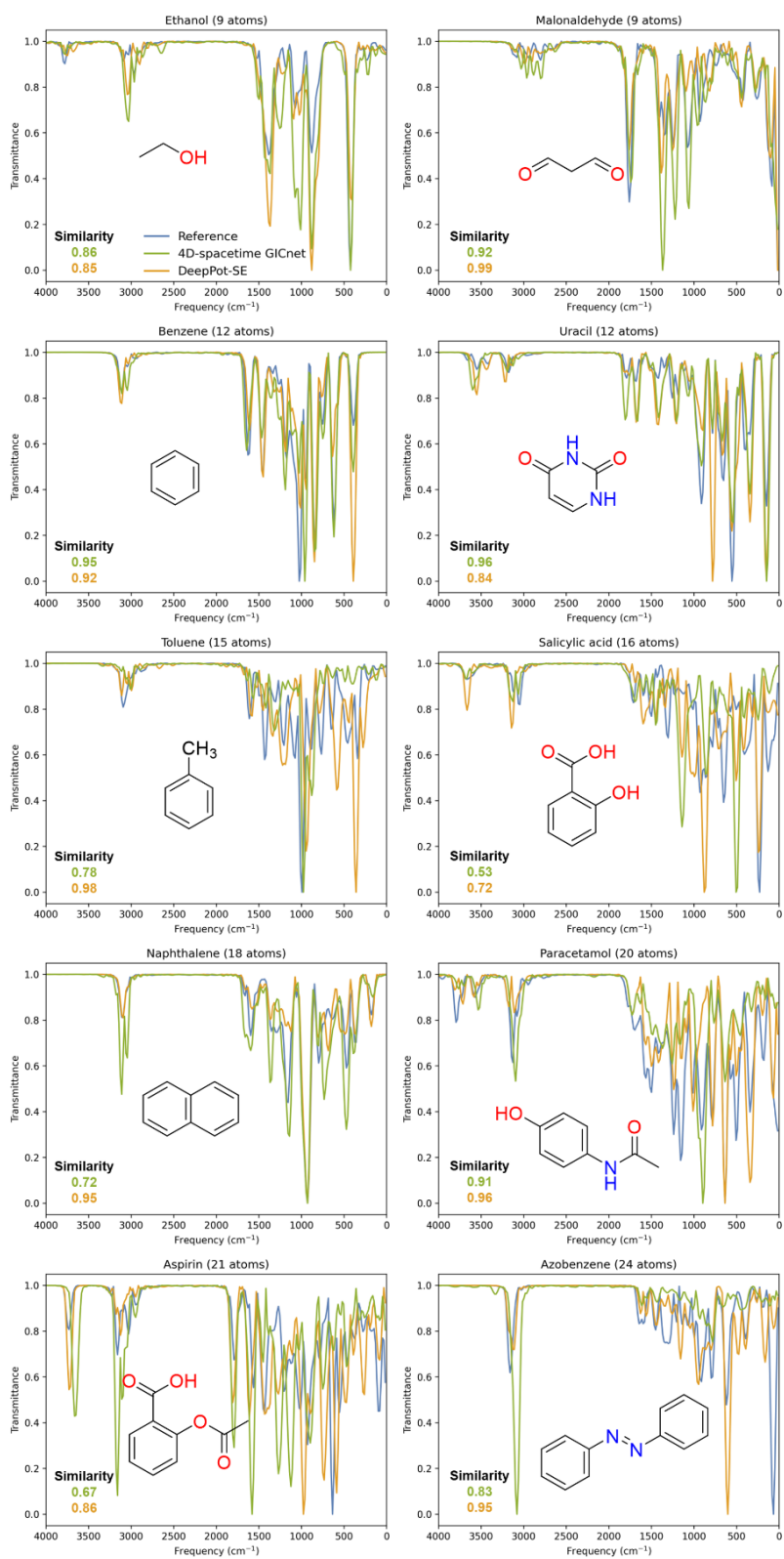

**Figure 6. Power spectra of the MD17 molecules.** Spectra obtained from molecular dynamics simulations with ANI-1ccx (reference, blue), 4D-spacetime GICnet (green), and DeepPot-SE (orange) are compared and similarities relative to the reference spectrum are calculated. For ethanol, malonaldehyde, benzene and paracetamol, MD was propagated up to 50 ps, for paracetamol up to 20 ps and for the remaining molecules – up to 10 ps. 20-fs models were used for benzene and uracil and 10-fs models for all other molecules.

Importantly, 4D-spacetime GICnet models allow us to generate long MD trajectories very fast with very high resolution (corresponding to the traditional MD time-step of 0.05 fs). In comparison, sequential traditional MD propagated even with the very fast 3D-MLP DeepPot-SE is substantially (ca. 10 times) slower (Figure 7). Thus, 4D-spacetime models can be used as a faster alternative to propagating trajectories with 3D-MLPs, which is quite impressive as the 3D-MLPs are already orders of magnitude faster than QM methods. In contrast to 3D-MLP, predictions with the GICnet within the segment can be very efficiently parallelized.

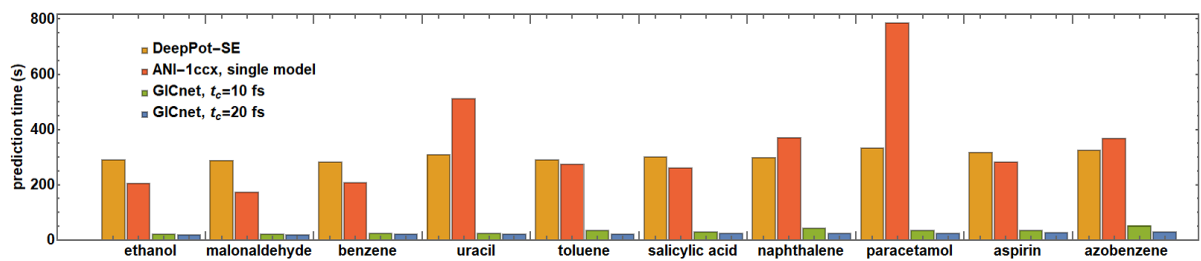

**Figure 7. 1-ps trajectory propagation time with GICnet, DeepPot-SE, and ANI-1ccx models.** Wall-clock times are measured on 18*Intel(R) Xeon(R) Gold 6240 CPU @ 2.60GHz. Note that for a fair comparison with DeepPot-SE, we show the timing of only one out of eight models in the ensemble for ANI-1ccx, but the actual reference calculations with the ensemble take even longer. In contrast, we report timing for GICnet dynamics based on an ensemble of eight models which also required evaluation of 3D-MLP energies.

**Insight into nuclear motions.** All nuclear motions in a molecule are subject to many-body quantum mechanical interactions and thus are coupled with each other to different extend. How they are exactly coupled is, however, difficult to quantify. The 4D-spacetime GICnet models offer unique insight into the coupled nuclear motions as we can interpret the NN predictions for each internal coordinate at a future time by analyzing the impact of initial values of all internal coordinates and their time derivatives. As an example, we analyze cis-trans isomerization trajectory of azobenzene with its GICnet model using the integrated gradients (IG) approach[54] which was also previously applied for quantitative interpretation of NN models for chemical reactions[55]. To assess how much nuclear motions and velocities contribute to the isomerization, we consider both forward and backward cis-trans isomerization process and calculate IG values for initial conditions 10 fs before the critical dihedral angle C–N=N–C reaches 90° (Figure 8).

The IG values are used to attribute the prediction of the critical dihedral angle to each of the inputs (see the SI). We can interpret IG value in terms of its magnitude and sign. Magnitude measures how strong the feature's impact is on the change of dihedral angle C–N=N–C, i.e., the cis-trans isomerization. It can be used to analyze the importance of features selected for the reaction and might also be useful for feature selection in the future. If the IG value is close to zero, it means this feature contributes little to the overall isomerization. A negative IG value indicates that the increase of this feature will decrease the dihedral angle and vice versa.

Time, the internal coordinate and velocity of the C–N=N–C angle have large IGs and the overall magnitudes of IG in dihedral angles are larger than those in bond lengths and bond angles. Many of the features have basically negligible effects on this critical dihedral angle. While these results are rather expected, the analysis of IG values also reveals much less obvious and surprisingly significant effects of other coordinates and time derivatives such as C2–C1–C3–C7, C6–C2–C1–C3, C13–C12–N24–N23 dihedral angles on one ring and C10–C6–C2–C1 on the other. We also observe noticeable differences in forward and backward propagations, e.g., the C21–C17–C13 bond angle has a noticeable effect, while in backward propagation, C17–C13–C12–N24 has overall the largest impact. The non-negligible effect is also of the C21–C17 bond length that highlights how complex and hard to interpretation are many-body quantum mechanical interactions.

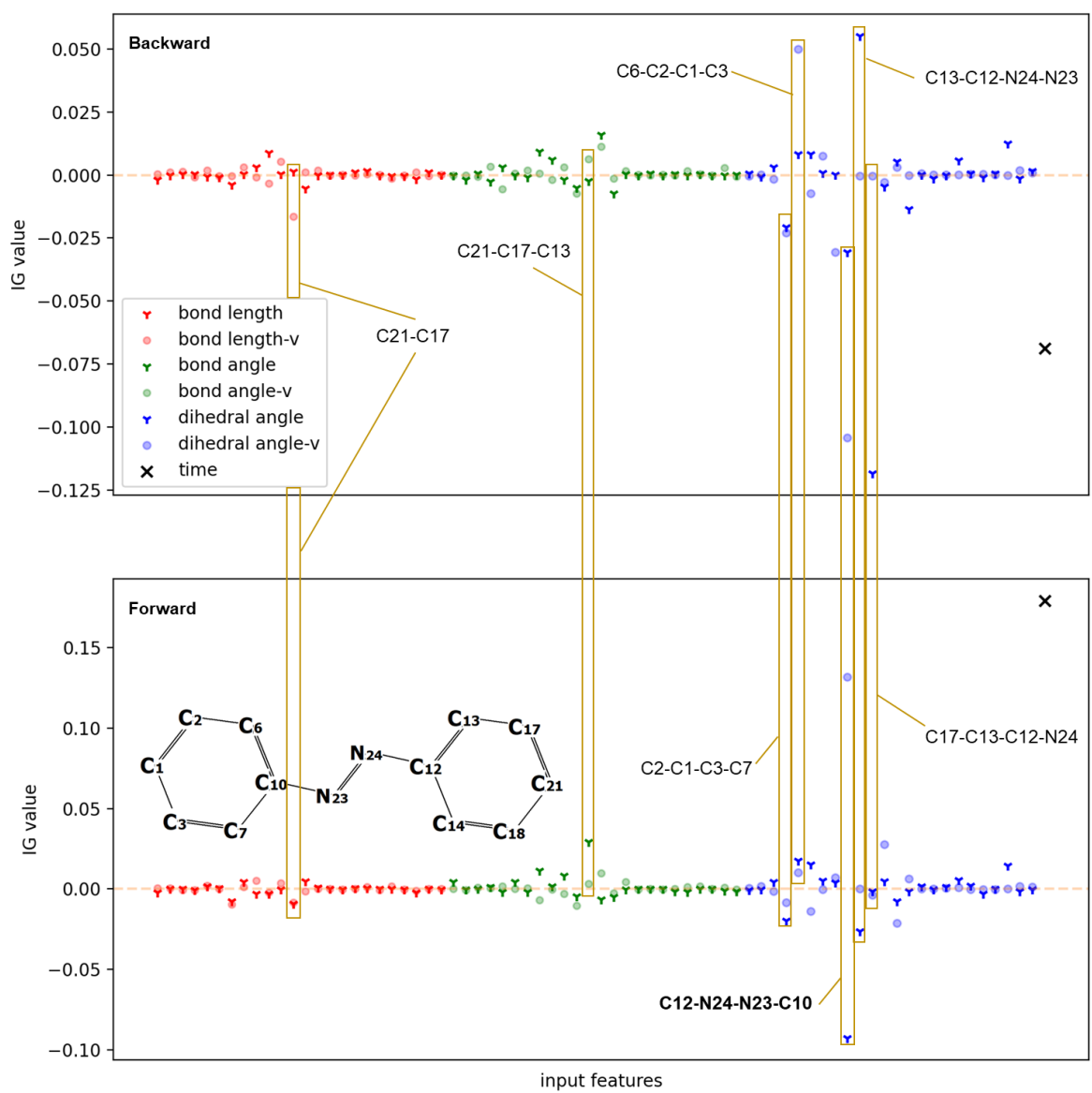

**Figure 8. Integrated gradients (IG) for cis-trans isomerization of azobenzene.** Values are calculated with the GICnet model of azobenzene for initial conditions in the forward and backward trajectories 10 fs before the dihedral angle C–N=N–C reaches 90°. Red points correspond to bond lengths, green – bond angles, and blue – to dihedral angles; their time derivatives are denoted with '-v'.

In summary, 4D-spacetime atomistic AI models proposed here have the potential to revolutionize the investigation of the atomistic dynamical processes, allowing for performing atomistic research in new ways, propagating very long-time trajectories with high QM-quality accuracy, achieving extensive statistical sampling, e.g., for detecting rare events, and finding reaction pathways. While here we have introduced the first models which focused on nuclear positions, analogous models may be in the future built for various properties of interest, such as molecular orbitals and dipole moments. Many issues remain to be addressed, such as testing and developing the methods for large atomistic systems such as biological macromolecules and solid-state materials, automatic construction of the data sets with active learning, and adapting the models for other types of simulations such as constant-temperature MD, which we are currently investigating.

**Data availability**

Data is reported in the manuscript and Supporting Information, models are available with the code; pre-trained models are available with the code. Any other data can be obtained from the corresponding author upon reasonable request.

**Code availability**

Code for 4D-spacetime models is implemented in the development version of the open-source MLatom program[56,57] and this version is publicly available from https://github.com/dralgroup/mlatom/tree/gicnet. The simulations can be performed online using the MLatom@XACS (http://XACScloud.com) cloud computing service. MLatom implementation of MD with 3D-MLPs is also publicly available as described in Ref. 46.

**Acknowledgments**

P.O.D. acknowledges funding by the National Natural Science Foundation of China (No. 22003051 and funding via the Outstanding Youth Scholars (Overseas, 2021) project), the Fundamental Research Funds for the Central Universities (No. 20720210092), and via the Lab project of the State Key Laboratory of Physical Chemistry of Solid Surfaces. This project is supported by Science and Technology Projects of Innovation Laboratory for Sciences and Technologies of Energy Materials of Fujian Province (IKKEM) (No: RD2022070103). The authors also thank our group member Peikun Zheng for stimulating discussions.

**Author contributions**

P.O.D. conceived the idea, designed and supervised the project, co-developed the methods, wrote the manuscript with input from all authors, and performed implementations and calculations for the $H_2$ system. F.G. co-developed the methods, performed the method implementation, model training, testing, and analysis. L.Z. performed dynamics simulations, model training, and analysis. Y.F.H. implemented 3D-MLP dynamics, power spectra simulations, and spectra similarity. Y.C. suggested and performed the analysis of insights. A.U.

assisted in the optimization of the neural network architecture. All authors discussed the results and revised the manuscript.

**Competing interests**

The authors declare no competing interests.

**Materials and correspondence**

Correspondence and requests for materials should be addressed to P.O.D.

**Supporting Information**

Additional results, discussion, analysis, and details about the methods as well as a video with molecular dynamics.

# Supplementary Information For Four-Dimensional-Spacetime Atomistic Artificial Intelligence Models

Fuchun Ge[1], Lina Zhang[1], Yi-Fan Hou[1], Yuxinxin Chen[1], Arif Ullah[2], Pavlo O. Dral[1*]

[1]*State Key Laboratory of Physical Chemistry of Solid Surfaces, College of Chemistry and Chemical Engineering, Fujian Provincial Key Laboratory of Theoretical and Computational Chemistry, and Innovation Laboratory for Sciences and Technologies of Energy Materials of Fujian Province (IKKEM), Xiamen University, Xiamen, Fujian 361005, China*

[2]*School of Physics and Optoelectronic Engineering, Anhui University, Hefei 230601, China*

E-Mail: dral@xmu.edu.cn

**Table of Content**

## Learning 4D Spacetime of a Hydrogen Molecule

4D-spacetime models which can make predictions for any time are conceivable for the simplest systems with very few degrees of freedom as we demonstrate for a hydrogen molecule $H_2$. If we take a typical hydrogen $H_2$ BOMD trajectory, the H–H bond length periodically oscillates with time. Thus, we can construct our first 4D-A$^2$I model $R_t = f(t)$ by using a simple periodic ML algorithm. We use a single set of initial conditions $\{R_0, v_0\}$ and just a handful of the training points within one period of BOMD MD trajectory propagated with the 3D-MLP surrogate model for the full configuration interaction QM approach (Figure S1a, see Supplementary Methods). This periodic 4D-spacetime model then can easily predict the bond length of $H_2$ at any time point in the future or in the past. Similarly, we have recently demonstrated that for some known asymptotic limits, ML can be trained to predict quantum dynamics up to the infinite future time, but this was done only for a model Hamiltonian and not for nuclear positions.[1] Predictions made with the 4D-A$^2$I model for $H_2$ are practically instantaneous while propagating sequential BOMD of even such a small system as $H_2$ up to 15 fs with 3D-MLP and a small time-step of 0.005 fs took longer than 30 min; propagating with full configuration interaction QM method would take several months on a desktop computer (see details below).

### *Computational Details And Methods For Hydrogen Molecule*

The BOMD trajectory was started with an initial bond length of 0.5 Å and zero velocity and propagated with 0.005 fs time step up to 15 fs with the NVE ensemble using Newton-X program[2, 3]. Forces required to propagate trajectory were calculated as the negative first derivatives of the potential energy with respect to the configuration of the geometry at a given time step. Using full configuration interaction (FCI) to generate forces would be too expensive, thus, we create a surrogate MLP model on 451 points along the potential energy curve of $H_2$ between 0.5 to 5 Å with energies calculated at FCI/aug-cc-pV6Z and taken from Ref. 4 (the open-source data can be downloaded from http://mlatom.com/aqctutorial/). For training MLP, we used kernel ridge regression with the Matérn kernel function ($n = 1$) and hyperparameters $\sigma = 96.5$ and $\lambda = 1.00 \cdot 10^{-13}$ as implemented in MLatom[5-7]. $\sigma$ and $\lambda$ were found by the grid search. We used a prior value equal to the mean of the energies in the data set (prior value is subtracted from the reference energies before the model is trained, i.e., the data is centered).

Periodic ML model was trained using the periodic kernel function as implemented in MLatom[5-7] (using the definition of sklearn[8]):

$$k(\mathbf{x}_i, \mathbf{x}') = \exp\left(-\frac{2}{\sigma^2}\sin^2\left[\frac{\pi}{p}\sqrt{\sum_{j}^{n}(x_{ij} - x'_j)^2}\right]\right), \quad (S1)$$

where $p$ is a period. The period in BOMD trajectory of H–H stretching was found to be equal to 9.565 fs. $\mathbf{x}_i$ is the feature vector for the $i$th training point, while $\mathbf{x}'$ is the feature vector for the point of interest. $n$ is the length of these vectors. Optimized hyperparameters were $\sigma$ = 0.741 and $\lambda$ = 5.30·10$^{-7}$.

The non-periodic ML model used for both recursive and non-recursive predictions was trained with the Gaussian kernel and optimized hypermeters $\sigma$ = 2.15 and $\lambda$ = 1.06·10$^{-9}$. The time-reversible model was the same, but with optimized hyperparameters $\sigma$ = 1.23 and $\lambda$ = 8.87·10$^{-11}$ and trained augmented data set, where for the artificially added points we subtracted period of 9.565 fs from all training times and used unchanged bond lengths from the trajectory.

All ML models trained on the trajectory used only 23 training points for time 0.000, 0.045, 0.145, 0.245, 0.395, 1.395, 2.420, 3.445, 4.445, 4.620, 4.720, 4.780, 4.840, 4.940, 5.115, 6.120, 7.120, 8.145, 9.165, 9.315, 9.415, 9.515, and 9.565 fs. Since we used the same initial condition, we can train ML as the function $R_t = f(t)$ rather than $R_t = f(R_0, v_0, t)$ as $R_0 = 0.5$ and $v_0 = 0$ are trivially the same in the training data. For recursive predictions, the bond length at the time of the end of the period (9.565 fs) is essentially the same as at the initial time and thus the model can make predictions starting from this point.

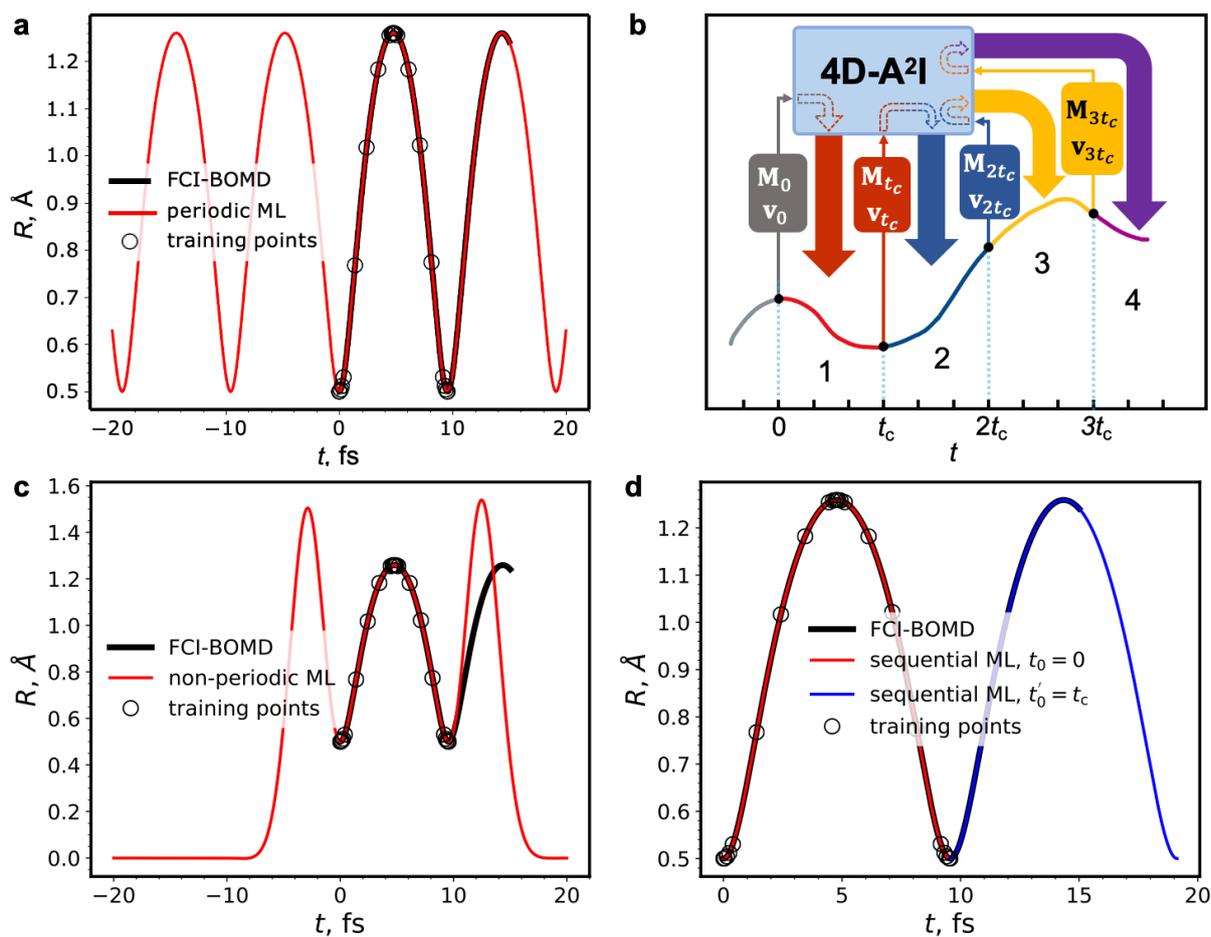

**Figure S1. 4D-spacetime trajectory prediction strategies on an example of the hydrogen molecule. a**. Predictions with periodic machine learning (ML) model (red) and reference full configuration interaction (FCI) Born–Oppenheimer molecular dynamics (BOMD) trajectory in black. **b**. Making predictions with a single 4D-spacetime atomistic artificial intelligence (4D-A$^2$I) model with a sequential strategy which makes prediction for any time in the next trajectory segment using initial conditions taken from the end of the previous trajectory segment. **M** – nuclear positions, **v** – nuclear velocities, the indices of **M** and **v** indicate the time step they are taken from, $t_c$ is time cutoff used in training. **c**. Non-periodic 4D-spacetime (ML) model (red) fails to make direct predictions beyond the training points. **d**. Non-periodic 4D-spacetime (ML) model (red) can make correct predictions in the future using a sequential strategy (shown two sequential steps, one in red with predictions starting from $t_0 = 0$ fs and blue with predictions starting from $t'_0 = t_c$; $t_c$ is time cutoff set equal to the time of one period).

The ultimate goal of a complete 4D-spacetime model is to describe an atomistic system across infinite 4D spacetime, i.e., predict dynamics trajectories for any combination of initial conditions and any time. While such models are conceivable for simplest systems with very few degrees of freedom as we have seen for $H_2$, with the increased number of atoms, building such models will become impossible due to the curse of dimensionality. A practical solution is to learn segments of trajectories of polyatomic systems up to some finite distant future within time cutoff $t_c$. A model trained within $t_c$ can be used to make predictions $f(\mathbf{M}_0, \mathbf{v}_0, t)$ only within the segment $[0, t_c]$ defined by a time cutoff, but we can extrapolate beyond this cutoff by making sequential predictions similar to traditional MD (Figure S1b). We start by predicting $\mathbf{M}_{t_c}$ and $\mathbf{v}_{t_c}$ with the model $f(\mathbf{M}_0, \mathbf{v}_0, t_c)$ for $\mathbf{M}_0$ and $\mathbf{v}_0$ (taken at the initial time $t_0 = 0$) and for time $t = t_c$ at the end of the segment. Then we can set new initial time $t'_0 = t_c$ and use $\mathbf{M}_{t_c}$ and $\mathbf{v}_{t_c}$ as new initial conditions to make predictions $f(\mathbf{M}_{t_c}, \mathbf{v}_{t_c}, t)$ with the same 4D-spaetime model and time $t$ again within the segment $[0, t_c]$ relative to $t'_0 = t_c$. The latter segment corresponds to the next segment in the trajectory $[t_c, 2t_c]$ relative to $t_0 = 0$; this process can be repeated as many times as required. The sequential approach also works for the above example of $H_2$, when a non-periodic 4D-spacetime model trained within a period can be successfully used (Figure S1d, see Supplementary Methods). Note that such a non-periodic model fails if we attempt to use it directly to make predictions $R_t = f(t)$ for a time beyond the time cutoff, i.e., for $t > t_c$ (Figure S1c, see Supplementary Methods). Our sequential approach is still much more computationally efficient than traditional 3D MD trajectory propagation because the latter would require integration during a large number (e.g., hundreds as we will see next for polyatomic molecule ethanol) of very small time steps to reach distant time $t_c$.

**Computational Details**

AIQM1[9] calculations were performed with the MLatom program interfaced to the MNDO program[10] (providing the ODM2*[11] part of AIQM1), TorchANI 2.2[12] (providing modified ANI-type[13] neural-network part of AIQM1), dftd4[14] (providing the D4-part[15] of AIQM1). ANI-1ccx[16] and DeepPot-SE[17] calculations were performed with MLatom interfaced to TorchANI 2.2 and DeePMD-kit v1.2.2[18], respectively.

**Supplementary Methods**

*GICnet Model*

The nuclear coordinates and their time derivatives are transformed from Cartesian coordinates into a modified version of internal coordinates (IC), which introduces 3 predefined dummy atoms whose position vectors are the 3 bases of the Cartesian coordinate system to keep the degrees of freedom unchanged (inspired by related approaches in MD using generalized ICs[19]). The choice of lengths, angles, and dihedrals used in IC was manually defined for each molecule.

The damping function used in the GICnet model (Eqs. 5 and 6 of the main text) is

$$f_{\text{damp}}(t) = 1 - e^{-(t/a)^2}. \tag{S2}$$

It is easy to see that $f_{\text{damp}}$ and its time derivative $df_{\text{damp}}(t)/dt$ at $t = 0$ are zero, i.e., $f_{\text{damp}}(0) = 0$ and $df_{\text{damp}}(t)/dt = 0$ as required. We used an adjustable parameter $a = 0.1$ fs.

The 4D-spacetime GICnet model is an ensemble of 8 individual neural networks (NNs). Each of them is a feed-forward, fully-connected $(6N_{\text{at}} + 1):1024:512:256:128:64:3N_{\text{at}}$ neural network (NN), where $N_{\text{at}}$ is the number of atoms in the molecule. The Gaussian error linear units (GELU)[20] were used as the activation functions. Models were trained with TensorFlow 2.4.0[21]. Nuclear positions ($\mathbf{R}_t$) were predicted by the output layer (i.e., using multi-output learning). Velocities $\mathbf{v}_t$ were estimated by using automatic differentiation as implemented in TensorFlow. Potential energies were predicted for a given 3D structure using an auxiliary 3D-MLP model represented by a feed-forward, fully-connected $3N_{\text{at}}:512:256:128:64:1$ NN with GELU (Figure 2 of the main text). All input was centered and normalized by the standard deviation around the mean.

The NN loss function in Eq. 7 of the main text is used for training the GICnet models includes terms with weighted mean root-mean-squared errors (RMSEs) in bond lengths, bond angles, dihedral angles, velocities of bond lengths, velocities of bond angles, velocities of dihedral angles, and potential energies. The mean RMSE is calculated by averaging the RMSEs in the features of the same type (i.e., length, angle, or dihedral angle).

The weights for different terms in Eq. 7 of the main text are adjusted manually to $w_B = 1.0$, $w_A = 2.0$, $w_D = 4.0$, $w_{vB} = 16.0$, $w_{vA} = 20.0$, $w_{vD} = 24.0$, $w_{E_{pot}} = 1.0$. The errors were also weighted by the corresponding time duration according to the formula $w(t) = e^{-t} + 1$ (smaller weights for more distant future). The training set for each molecule contained 1 million points and a small validation set of 1024 points was used to ensure there is no overfitting. During training we saved models with the best validation error, training was performed using Adam optimizer[22].

The auxiliary 3D-MLP models were trained on 1 million points with additional 1024 points for validation using the Adam optimizer. The input was centered, normalized, and used the same internal coordinates as for the GICnet model; in addition, labels were also centered and normalized.

Neither the GICnet nor the auxiliary 3D-MLP model is permutationally invariant by construction, i.e., permuting the order of homonuclear atoms in the input structure is not ensured by construction to produce the chemically equivalent output. This might be a problem for the training data sets not containing sufficient information about the permutations as is known in the field of the machine learning potentials.[23, 24] However, it is also known from the same field, that for our test molecules, even data sets with thousands of points are sufficient to contain enough information about permutations so that even non-permutationally invariant models can produce as accurate results as models which are explicitly constructed to be permutationally invariant.[25] In addition, for very small data sets, if required, it is possible to include the permutational information in the training set by artificially generating additional structures, i.e., exploiting data augmentation which is standard in the field of ML and is also sometimes used in the field of MLPs[23, 26].

*Prediction*

The MD trajectories using the GICnet model are propagated in a segment-by-segment manner, which differs from the step-by-step traditional MD methods. The length of a segment is defined by segmentation time $t_{\text{segm}}$, which should be no larger than the $t_c$ used in model training. In each segment, predictions from all NNs were made with the initial conditions and the $t_{\text{segm}}$ first.

For a single NN, the prediction is made with the following steps: remove linear and angular momenta, transform initial conditions from Cartesian coordinates to IC, center and normalize, evaluate NN to get $\mathbf{R}_t$, run automatic differentiation to get $\mathbf{v}_t$, obtain and back convert $\mathbf{R}_t$ and $\mathbf{v}_t$ from IC to Cartesian coordinates, remove linear and angular momenta again.

After the evaluations of potential and kinetic energies, the NN with the least total energy shift is chosen to make the predictions for this segment in the requested temporal resolution. Then the best prediction at $t_{\text{segm}}$ becomes the initial conditions for the next segment.

To enhance the stability of the MD prediction, the best model's predictions of a segment go into the buffer, if the 1 kcal/mol threshold of the total energy shift cannot be met in the next segment, the propagation will be rolled back to this point and the second-best model will make the predictions for the segment instead. If it failed again, the third-best model will be chosen, and so forth, until no model is left or no model gives a prediction that meets the threshold. Then the propagation will be rolled back to the previous segment to repeat this process until the buffer of 10 segments runs out. Once the buffer has been run out, the threshold will be temporarily increased by a factor of 2 until the buffer is filled up again.

In addition, we have observed a deterioration in the stability of predictions after long-time propagations. This issue might arise because the definition of the redundant internal coordinates is not unique relative to the dummy atoms and, hence, the long-time dynamics may lead to the gradual change of molecular orientation defined by internal coordinates relative to the dummy atoms that goes into an uncovered region of the training data. We solved this problem by resetting internal coordinates to the values in initial conditions when the buffer is exhausted. This transformation does not change the geometry but effectively reorients the molecule in the XYZ space.

*Reference Trajectory Generation*

For each molecule except for azobenzene and ethanol from the MD17 dataset[24] (benzene, uracil, naphthalene, aspirin, salicylic acid, malonaldehyde, toluene, and paracetamol), 31 reference BOMD trajectories are propagated using MD implementation[27] in MLatom. For ethanol, 34 reference BOMD trajectories are propagated. Initial conditions are generated by a Wigner distribution[28, 29] in the ground vibrational state of the equilibrium geometry optimized by ANI-1ccx[16] level with MLatom under 0 K temperature. The distribution is created using normal modes obtained at the ANI-1ccx level. The energies and energy gradients at each classical time step are also computed at the ANI-1ccx level. The time step of the classical trajectories is set to be 0.05 fs and the maximum run time is 10 ps (if not mentioned otherwise). Trajectories are propagated with the NVE ensemble, and the linear and angular momenta of initial conditions are removed before propagation.

For azobenzene, which experiences *cis*-to-*trans* thermal isomerization, 60 reference BOMD trajectories are propagated starting from initial conditions generated around six different geometric regions and each type of these initial conditions leads to 10 trajectories. The six different geometric regions are around six structures taken from the literature[30] including *cis*-azobenzene and *trans*-azobenzene equilibrium structures as well as the energy maxima structures in four constructed *cis*-to-*trans* thermal isomerization transition paths – rotation, inversion, rotation assisted by inversion (there are two local maxima geometries and the first one is taken) and "optimized" rotation paths. Initial conditions around *cis*-azobenzene and *trans*-azobenzene equilibrium regions are generated via Wigner distribution under 0 K temperature in the ground vibrational state of *cis*- and *trans*- azobenzene equilibrium geometry with distribution created using normal modes obtained at the AIQM1[9] level. For geometric regions near energy maxima in four constructed *cis*-to-*trans* thermal isomerization transition paths, Gaussian-distributed random velocities are generated under 0 K temperature for four specific geometries (energy maxima structure in rotation, inversion, rotation assisted by inversion and "optimized" rotation paths) using NEWTON-X[31] by setting kinetic energy to 3.17, 3.70, 3.71 and 3.25 eV, respectively. Kinetic energies are chosen according to the differences between the sum of electronic and zero-point energies of the *trans*-azobenzene equilibrium structure and the electronic energies of energy maxima structures. After initial conditions are generated, the linear and angular momenta are removed and 60 trajectories of the azobenzene molecule are propagated in the same way as trajectories of other molecules are propagated, i.e., with energies and energy gradients computed at the ANI-1ccx level and the time step and the maximum run time set to be 0.05 fs and 10 ps, respectively.

Before training the 4D-spacetime GICnet and DeepPot-SE models, reference BOMD trajectories are checked and trajectories leading to explosions are removed. Thus, the number of the reference BOMD trajectories are 31, 31, 31, 31, 31, 21, 30, 17, 34, 59 for benzene, uracil, naphthalene, aspirin, salicylic acid, malonaldehyde, toluene, paracetamol, ethanol, and azobenzene, respectively. Since for paracetamol, only 17 trajectories were successful, we propagated additional 29 trajectories and we truncated 20 of them before they led to significant dissociation (RMSD relative to the first geometry greater than 5). For each molecule, one non-truncated reference trajectory is saved as the test trajectory, and other trajectories are used as training trajectories.

For each molecule, all trajectories except for one were used to generate the labeled training data. The unused trajectories were used as the test trajectories for evaluating the GICnet models.

*Test Errors*

The performance of the GICnet models was examined by applying them to the test data sets generated from the trajectories unused in training. For each molecule, segments are picked from the trajectory to form a 10000-point test set. Then the mean RMSEs are calculated as in the loss function of GICnet; the RMSEs were not weighted.

*RMSD Calculation*

The root-mean-square deviation (RMSD) value between the two geometries reported here is calculated in the following steps. First, each geometry is centered on the origin by its center of mass. Second, geometries are aligned using the rotation matrix calculated with the Kabsch algorithm implemented in the Python module RMSD (http://github.com/charnley/rmsd). Third, the RMSD is calculated between the Cartesian coordinates of two centered and aligned geometries.

*Integrated Gradients*

Integrated gradients[32] (IGs) are obtained by summing up gradients at all points along the straight path from the baseline to the input. Given a function $F: R^n \to [0, 1]$ and the interpolation constant $\alpha$ to perturb features, the IG of $i^{th}$ feature for baseline $\mathbf{x}'$ and input vector $\mathbf{x}$ is calculated as follows:[32]

$$IG_i(x) = (x_i - x_i') \times \int_{\alpha=0}^{1} \frac{\partial F(\mathbf{x}' + \alpha \times (\mathbf{x} - \mathbf{x}'))}{\partial x_i} d\alpha. \tag{S3}$$

In our case, we define baseline as the equilibrium structure of *trans*-azobenzene with zero velocity for each internal coordinate. Gradients of each input feature are computed in terms of the dihedral angle C–N=N–C. For analysis, we took the GICnet model trained with $t_c = 10$ fs.

**Supplementary Analysis**

*Estimate of Total Energy*

For ensuring the stability of MD propagation with the GICnet model we only estimate the end-of-the-segment total energy as described in the main text. These estimates are plotted for the test 10-ps trajectories in the figures below for all ten molecules (Figure S2. and Figure S3.). Results show that all molecules have either very small fluctuations throughout the trajectory or large but random fluctuations which rapidly return to the values around the initial value.

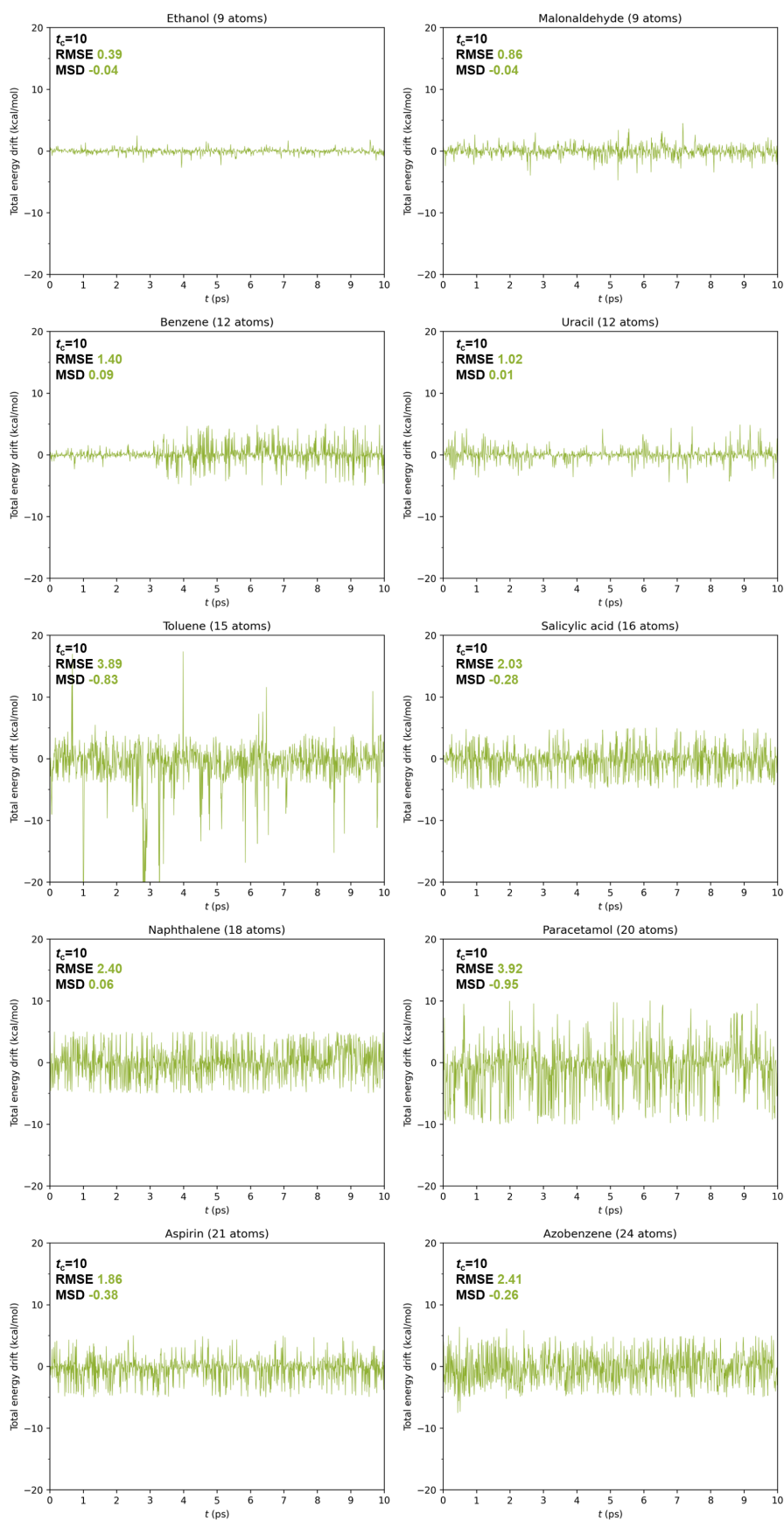

**Figure S2.** Estimated end-of-the-segment total energies in the test GICnet MD trajectories for all molecules and $t_c = 10$ fs is used. Energies are relative to the initial energy. RMSE – root-squared-deviation and MSD – mean signed deviation in kcal/mol.

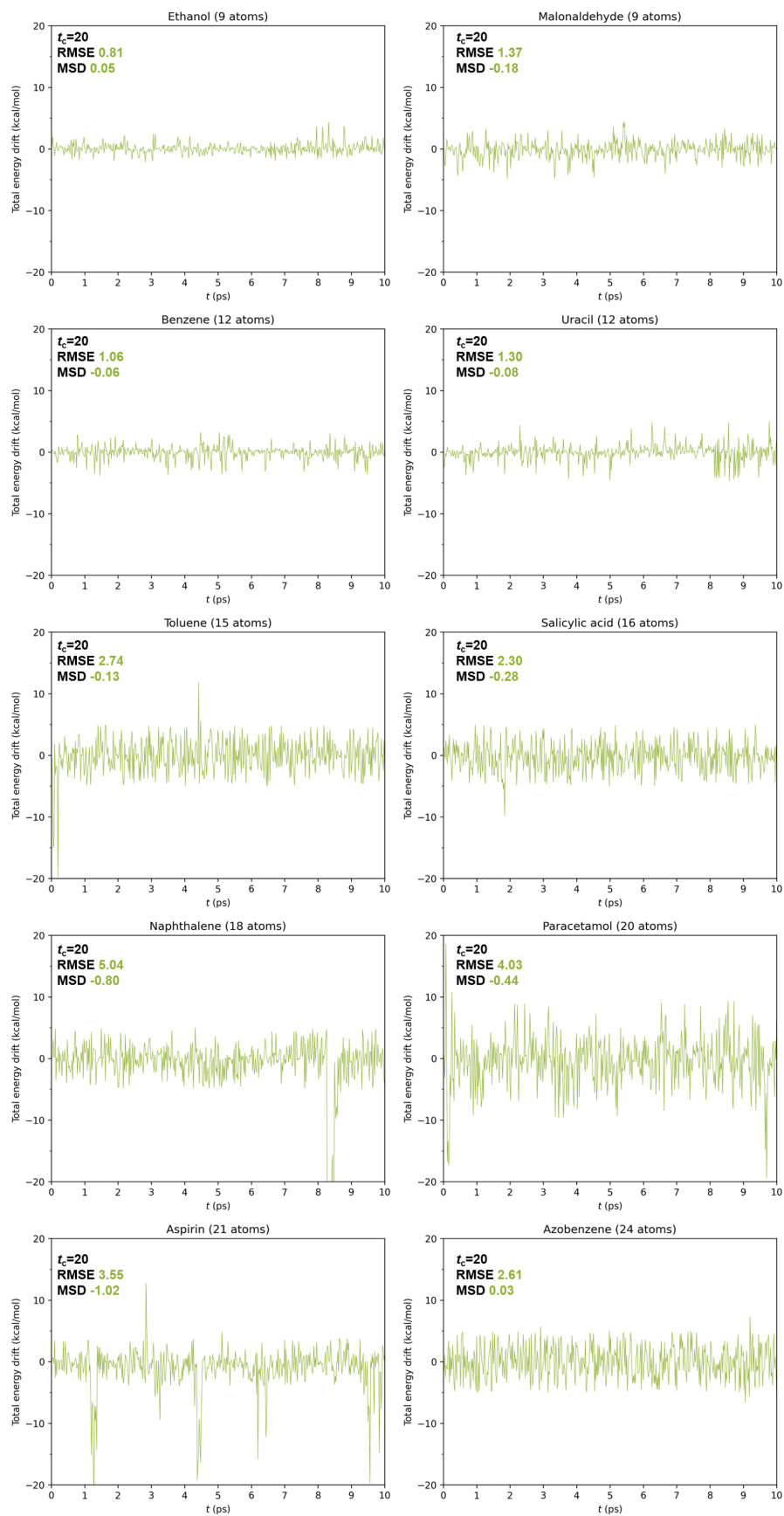

**Figure S3.** Estimated end-of-the-segment total energies in the test GICnet MD trajectories for all molecules and $t_c$ = 20 fs is used. Energies are relative to the initial energy. RMSE – root-squared-deviation and MSD – mean signed deviation in kcal/mol.

*Spectrum Convergence*

We propagated the test trajectory and generated the power spectrum for each molecule. Then we calculated the similarity of this power spectrum to power spectra calculated from trajectory portions of different lengths. In cases of ethanol, malonaldehyde, benzene, and paracetamol, we found that 50 ps was sufficient to converge the GICnet spectra while for paracetamol we needed 20 ps and for other molecules 10 ps was sufficient (Figure S4). The time step is set to be 0.05 fs with the exception of ethanol, where we used a 0.5 fs time step.

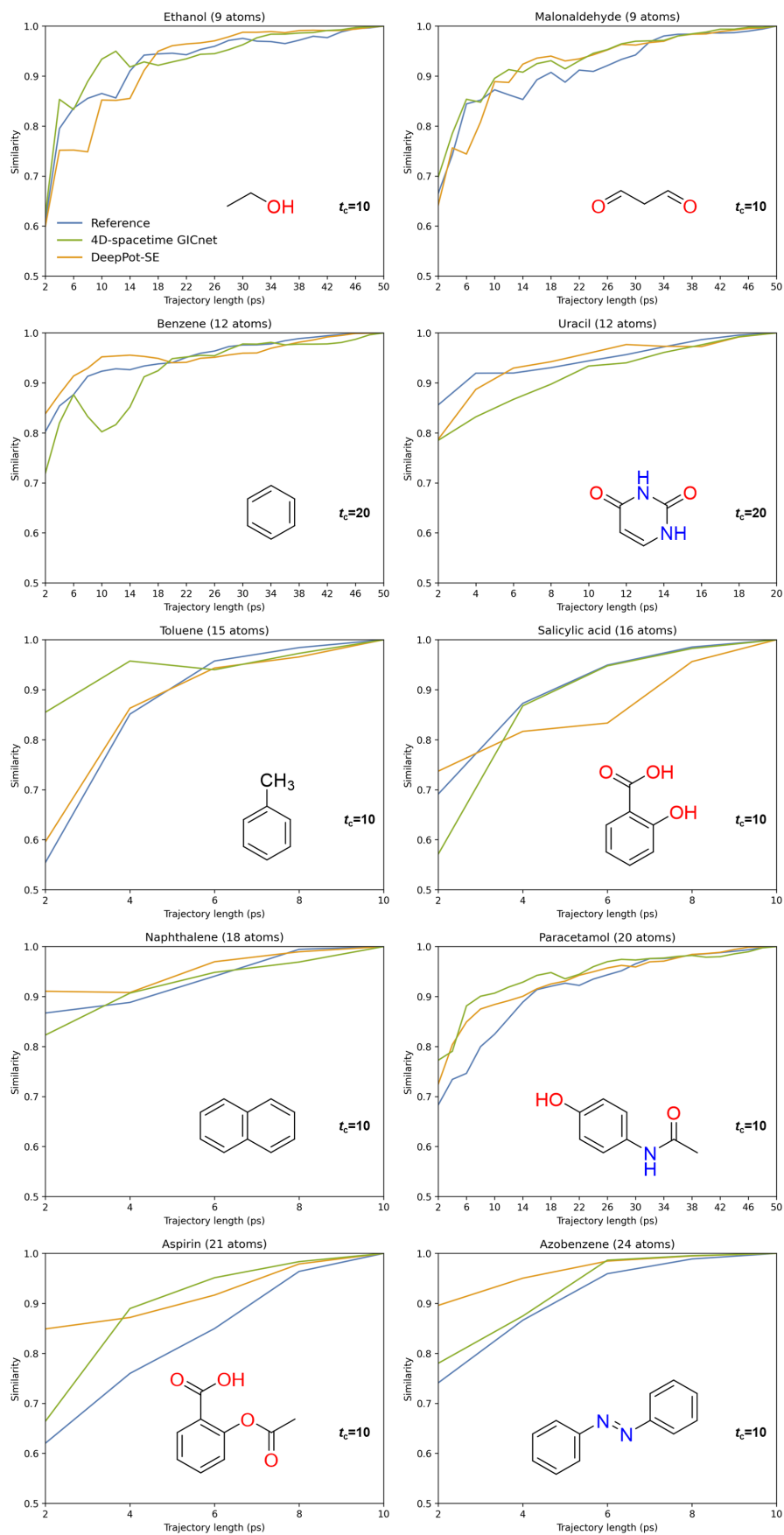

**Figure S4.** The similarity of spectra generated from trajectories of different lengths to the trajectory generated with the maximum simulation time (maximum value in the x-axis).

*Effect of Time Cutoff on the Power Spectra of Ethanol*

To investigate the impact of the choice of the time cutoff ($t_c$) in training data generation on the long dynamics, we trained the GICnet models with different choices of $t_c$ (1, 5, 10, 20, 30, and 50 fs) for the ethanol molecule, and ran 50-ps dynamics with them. The dynamics failed with models that have $t_c$ longer than 20 fs. The power spectra generated from the successful trajectories along with the reference ANI-1ccx and another machine learning potential DeepPot-SE spectra are shown in the Figure S5 below.

From the figure, we can see that all successful models gave visually reasonable spectra compared to the reference, while the 10 and 20 fs model gave the best similarities. Considering longer time cutoff models failed to complete the 50-ps dynamics, this result might indicate a ~5–20 fs cutoff to be the sweet point of the current model architecture for the molecule. The bad similarity with the smallest time cutoff of $t_c = 1$ fs indicates that the parameters of the model such as those in the dumping function may be non-optimal for this short cutoff.

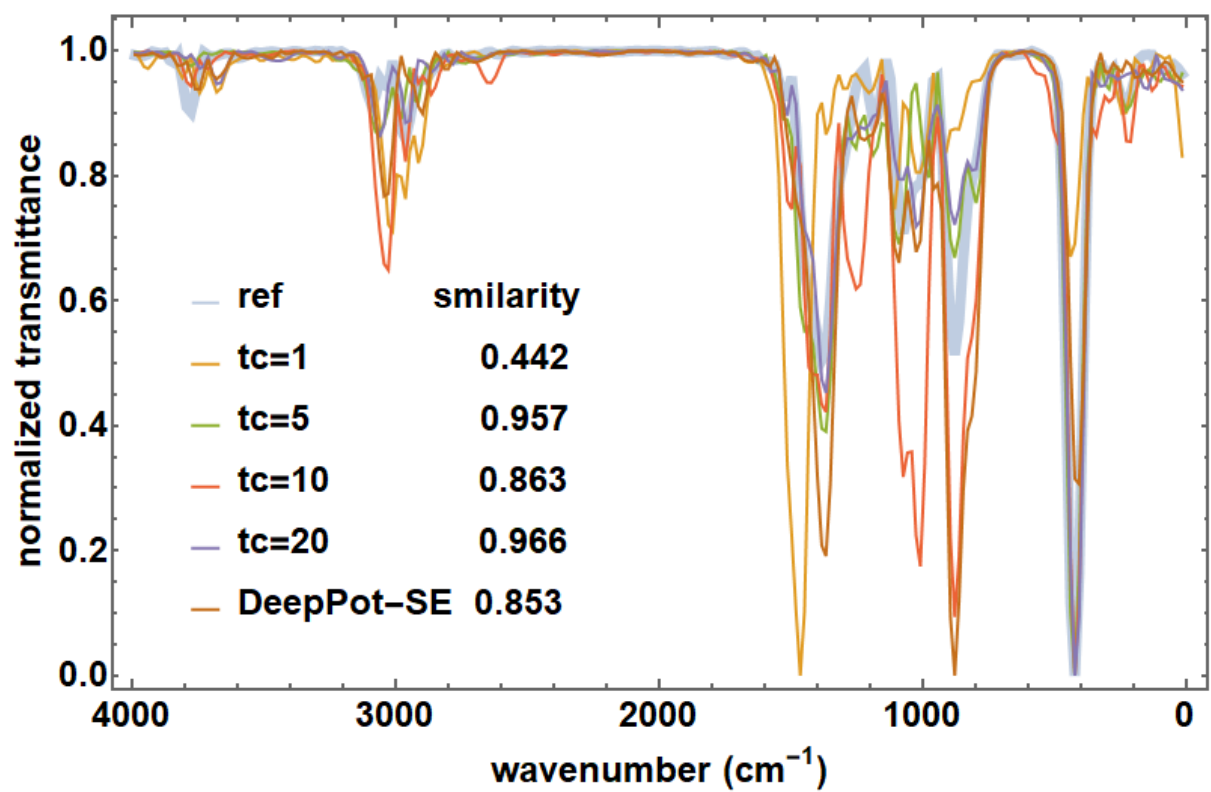

**Figure S5.** The power spectra and similarity score of the power spectra from 50-ps trajectories propagated with the GICnet model trained on different time cutoffs $t_c$ and with DeepPot-SE to the reference (ANI-1ccx) spectrum for ethanol.

*10- and 20-fs Power Spectra*

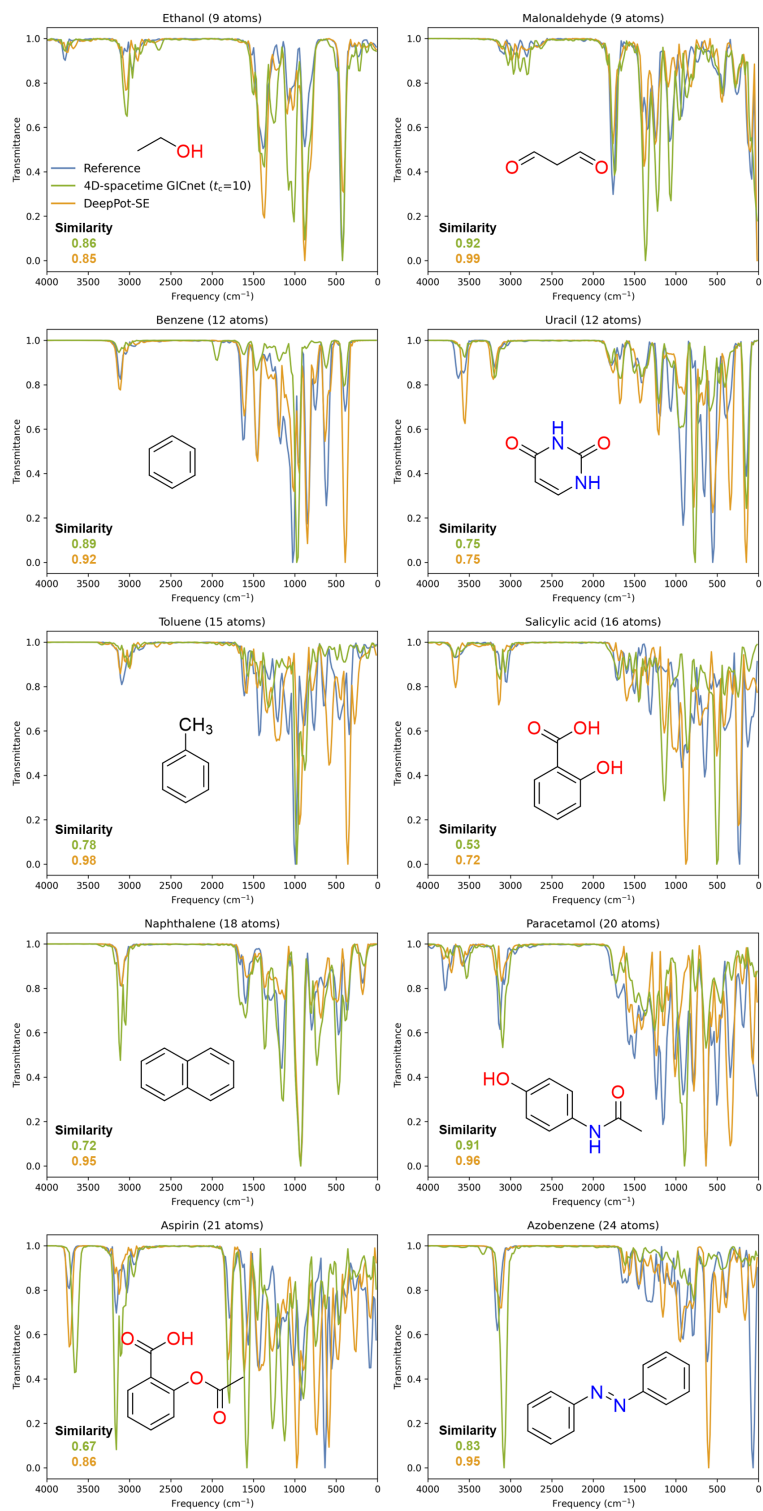

**Figure S6. Power spectra of the MD17 molecules.** Spectra obtained from molecular dynamics simulations with ANI-1ccx (reference, blue), 4D-spacetime GICnet (green, $t_c$ = 10 fs), and DeepPot-SE (orange) are compared and similarities relative to the reference spectrum are calculated. For ethanol, malonaldehyde, benzene and paracetamol, MD was propagated up to 50 ps, for the remaining molecules – up to 10 ps.

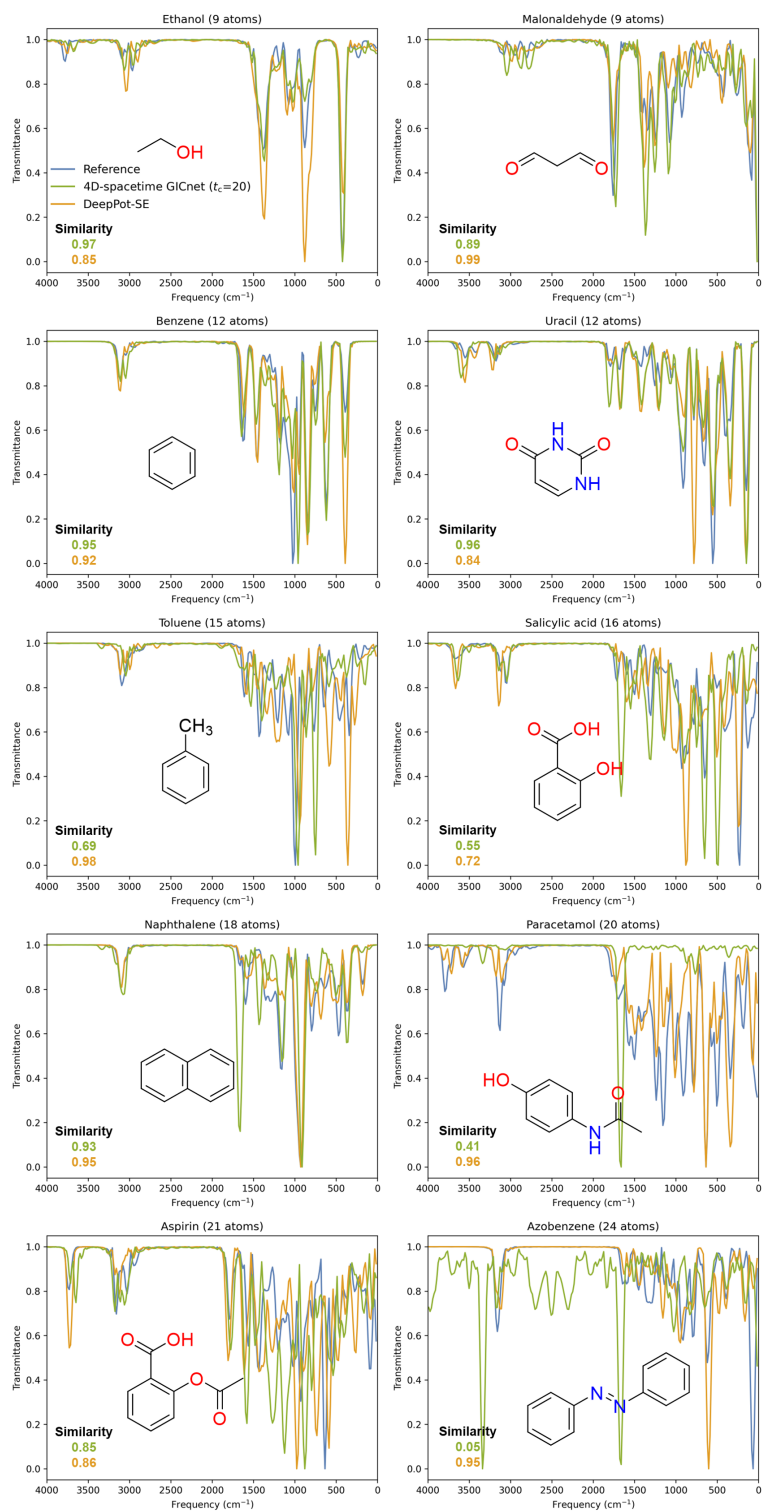

**Figure S7. Power spectra of the MD17 molecules.** Spectra obtained from molecular dynamics simulations with ANI-1ccx (reference, blue), 4D-spacetime GICnet (green, $t_c$ = 20 fs), and DeepPot-SE (orange) are compared and similarities relative to the reference spectrum are calculated. For ethanol, malonaldehyde, benzene and paracetamol, MD was propagated up to 50 ps, for uracil – up to 20 ps, and for the remaining molecules – up to 10 ps.

## Supplementary References